# Development of novel algorithm to visualize blood vessels on 3D ultrasound images during liver surgery

Fatemeh Salehihafshejani[1,*]. Alireza Ahmadian[2], Afshin Shoeibi[3], Roohallah Alizadehsani[4], Habibollah Dashti[5], Niloofar Ayoobi Yazdi[6], Abbas Khosravi[4], Saeid Nahavandi[4]

[1] Department of Medical Physics and Biomedical Engineering, School of Medicine, Tehran University of Medical Sciences, Tehran, Iran. Fatemeh.salehi.h@gmail.com
[2] Department of Medical Physics and Biomedical Engineering, School of Medicine, Tehran University of Medical Sciences, Tehran, Iran.
[3] Computer Engineering Department, Ferdowsi University of Mashhad, Mashhad, Iran
[4] Institute for Intelligent Systems Research and Innovation (IISRI), Deakin University, Victoria 3217, Australia

**Abstract**

Volume visualization is a method that displays three-dimensional (3D) data in two-dimensional (2D) space. Using 3D datasets instead of 2D traditional images improves the visualization of anatomical structures, and volume visualization helps radiologists and surgeons to review large datasets comprehensively so that diagnosis and treatment can be enhanced. In liver surgery, blood vessel detection is important. Liver vessels have various shapes and due to the presence of noise in the ultrasound images, they can be confused with noise. Suboptimal images can sometimes lead to surgical errors where the surgeon may cut the blood vessel in error. The ultrasound system is versatile and portable and has the advantage of being able to be used in the operating theatre. Due to the nature of B-mode ultrasound, 1-D transfer function volume visualization of images cannot abrogate shadow artifacts. While multi-dimensional transfer function improves the ability to define features of interest, the high dimensionality in the parameter domain renders it unwieldy and difficult for clinicians to work with. To overcome these limitations, an algorithm for volume visualization that can provide effective 3D visualization of noisy B-mode ultrasound images, which can be useful for clinicians, is proposed. We propose a method that is appropriate for liver ultrasound images focusing on vessels and tumors (if present) in order to delineate their structure and positions clearly to preempt surgical error during operation. This method can prevent possible errors during liver surgery by providing more detailed high quality 3D images for clinicians.

**Key Words:** Visualization, 3D ultrasound image, Volume Rendering, Liver surgery, Liver vessels.

## 1. Introduction

The liver is one of the most important intra-abdominal organs. Hepatitis, cirrhosis, and a variety of tumors are diseases that may occur in the liver [1, 2]. Liver biopsy is sometimes performed to obtain liver tissue for analysis in order to diagnose specific pathologies in the liver. In patients undergoing surgery, it may be possible to sample the tissue or remove it from the body during surgery. During surgery, unhealthy tissue is completely excised with least damage to the healthy tissue and vascular system of the liver [3]. These surgeries can be performed under ultrasound guidance in the operating theatre. Anatomy of liver and the vessel distributed is shown in Figure 1. Ultrasound is one of the most common methods used for imaging the liver as it is versatile, portable, non-invasive and low-cost. Ultrasound has the ability to detect changes during surgery in real-time. It does not involve nephrogenic contrast agents or ionizing radiation, hence it is well-suited for operating theatre use [4]. In liver surgery, because of deformation and organ shift during operation, the availability of ultrasound in the operation room is obligatory, and real-time ultrasound imaging is frequently used [5].

However, ultrasound images are limited by acoustic window access during acquisition; and image artefacts ("noisy" ultrasound images) during the interpretation. In liver surgery, blood vessel detection is important. Vessels have various shapes and because of noise in ultrasound images, they may be mistaken for noise [6]. As the images can sometimes be blurry, it is possible for surgeons to cut blood vessels in error. Thus, effective visualization is very important during liver surgery.

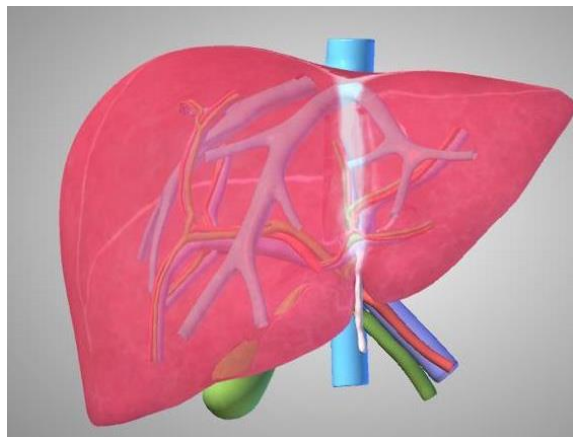

Figure 1. Schematic vascular system of liver.

Volume visualization or volume rendering is a method that displays three-dimensional (3D) medical images in a two-dimensional (2D) space. In recent years, with increasing availability of 3D image datasets such as computed tomography (CT), magnetic resonance imaging (MRI), and ultrasound, visualization methods have increased [7]. Using 3D datasets instead of 2D traditional images can improve visualization of anatomical structures. Using volume visualization helps radiologists and surgeons to review large datasets comprehensively, so that disease diagnosis and treatment can be improved [8]. This has been a popular tool for CT and MRI but not ultrasound because volume visualization in ultrasound imaging suffers from shadow artifacts in the images and lacks effective and usable algorithm for ultrasound visualization. Hence, the clinicians tend to use 2D images predominantly rather than 3D images [9].

Direct volume visualization (DVV), which maps volume points to composited image points directly, is one of the primary methods in volume visualization that is used in 3D medical images. The pipeline of DVV is shown in figure 2. The classification is mapping of sample data values to optical properties (i.e., opacity, color, etc.) performed by TF [10].

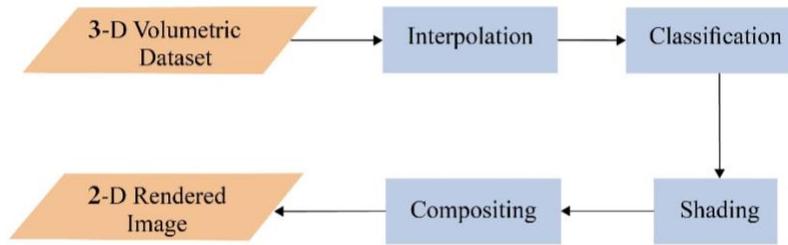

Figure 2. Pipeline of direct volume visualization.

1D-transfer function (TF) yields effective image results from tomographic reconstruction, such as in MRI and CT. Due to the nature of B-mode ultrasound, 1-D TF volume visualization of images cannot abrogate shadow artifacts as shown in figure 3 [11]. While multi-dimensional transfer function (MDTF) improves the ability to define features of interest, the high dimensionality in the parameter domain renders it unwieldy and difficult for clinicians to work with [11]. Hence, an effective algorithm for visualization of ultrasound images remains challenging despite advances in 3D ultrasound systems.

Despite recent research and advances in volume visualization, many clinicians demur using 3D images due to the limitations of the TF method and prefer to use 2D ultrasound images [11]. In this paper, we propose a method that is appropriate for the characteristics of ultrasound images, and we choose to focus on vessels and tumors (if present) to delineate them for the surgeon to avoid error during operation. In this method, previous TF is not used for classification. Each voxel corresponding to the region (i.e., vessel, tumor, etc.) it belongs to has its own opacity. As such, this method does not have the limitations of TF and retains interaction with the user. The schematic diagram of our work is shown in figure 4.

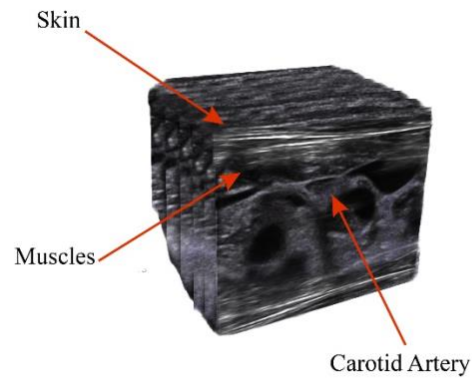

*Figure 3. Shadow artifact in traditional methods*

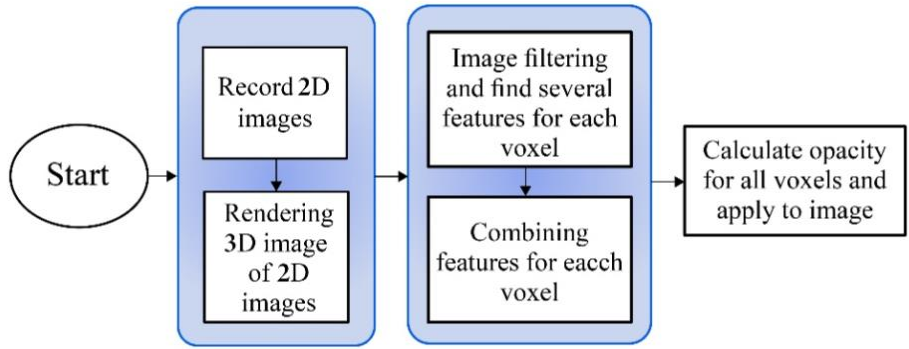

Figure 4. Schematic diagram of the steps of our work. The 2D images are acquired. Then for all voxels, global and local features are calculated to obtain their own opacity that applies to the image. In the end, the 3D image is rendered with these opacities.

## 2. Related Works

Various methods have been proposed for tumor segmentation in ultrasound images. In 1998, Yoshida presented a method to segment the objects in noisy space [12]. In 2006, Boen et al. developed seeded region growing method which represents the region of the user's target (such as a tumor) in a closed area [13]. In 2013, Kumar et al. presented an automated segmentation method that separates normal and tumor areas using statistical methods [14]. Egger et al. suggested a method for real-time segmentation of the tumor, which is capable of simultaneous feedback from the user that previous methods could not offer [15]. The visualization techniques used in the medical images are given below:

Focus-context visualization: This method highlights the focus points while maintaining the context to maintain the user's spatial and strategic visualization. There are several ways to do this and their difference is in the graphical design used in these methods [10]. This method is one of the best ways to solve the problem of shadow artifact that combines standard volumetric methods with the schematic. It highlights areas that are important for the user and less important areas are also shown in the spatial display [9]. This method is not based on TF. In 2006, Kruger et al. introduced a clear view method that uses curves and distance information [16].

In 2010, Wan et al. introduced transparency modulation, which proposed to visualize a schematic by using maximum intensity difference accumulation (MIDA) method with points of interest without the need for a TF [17].

Degree of importance is an important parameter used in most of the computer graphics displays [10]. For example in the wrong method, the color scheme of each object is considered to be important [18].

TF method: This method has many different types. It can be one-dimensional when only one variable is used to extract the property such as optical property. It can also be multidimensional and use other features such as gradient size and second derivative [19-21]. All the mentioned methods for displaying medical images are stated. But they have not specifically addressed the challenges of ultrasound.

In 2001, Fattal et al. developed a method for showing smooth surfaces in ultrasound that only represents one surface and does not represent different layers [22]. Mann introduced a volumetric ultrasonic system that increases the intensity of mode-B ultrasonic with elasticity information using a two-dimensional transmission function [23]. In 2008, Salzbrunn et al. proposed flow visualization method using a streamline predicate to combine flow topology with extracted features [24]. In 2013, Born et al. developed a method called line predicate to display blood flow. In this way, the blood flow structure is selectively shown with respect to flow rate, maximum velocity, etc. [25].

In the year 2014, Berge et al. [9] proposed an algorithm for 3D visualization of ultrasound images that extended the classic pipeline towards the focus-and-context algorithm. They introduced the point predicates for volumetric image data that proposed both local features (intensity and gradient) and global features (anatomical models at each sampling point), but did not focus on liver vessels.

In order to overcome the above mentioned shortcomings, an algorithm for volume visualization of liver for effective visualization and easy usage for clinicians in noisy B-mode environment is proposed. It can help considerably in liver operations and surgeons can use 3D images instead of traditional 2D images to get more detailed information. Therefore, it may reduce surgeon's need for radiologist in operating theatre.

None of these procedures have been specifically performed in the liver. They are incomplete and have not addressed the current challenges in the liver diagnosis. Currently, clinics in liver surgery depend on the expertise of radiologists with experience in Doppler or contrast agents in ultrasound [26]. Our goal is to present a method to visualize vessels and tumors in the liver with higher resolution and reduce the dependence of the user.

## 3. Material and Methods

The first step in this study is to acquire 3D ultrasound images of the liver. Many techniques have been developed for 3D ultrasonic imaging [27]. Depending on the methods used, they can be divided into *four* categories: 2D converter arrays, mechanical scanners, free-hand methods with spatial information, and non-spatial methods with no tools. Ultrasound systems using two-dimensional converter arrays can provide real-time 3D images of a given volume, but they are expensive and not readily available. Also, the output of most of the existing systems is 3D images designed solely for the display to the physicians. Systems that have probes with one-dimensional converter arrays are commonly used in most hospitals to record two-dimensional images. To use the usual two-dimensional systems and convert it into a 3D image, a tool is needed to render these 3D images in such a way as to render and reproduce 3D images.

### 3.1. Dataset

In this study, the ClarUs EXT-1M and probe with one-dimensional array C7-3R50NI-5 were used to record images using the Telemed ultrasound system (a PC-based system) (Figure 5). The method we used for 3D rendering in this system is 3DView software [28]. The 3DView is an optional module provided for 3D image and volume rendering. This software allows the information to be collected from different vertical, horizontal and, diagonal angles. Once the scan is complete, the data can be viewed in 3D from different angles. The free-hand 3D imaging of anatomical structures is performed with standard converters. This software complies with MDD 93/42 / EEC requirements for three-dimensional collection and reconstruction, which is one of the latest medical diagnosis requirements. The digitized images are put together and displayed in 3D using a special technique as shown in Figure 6.

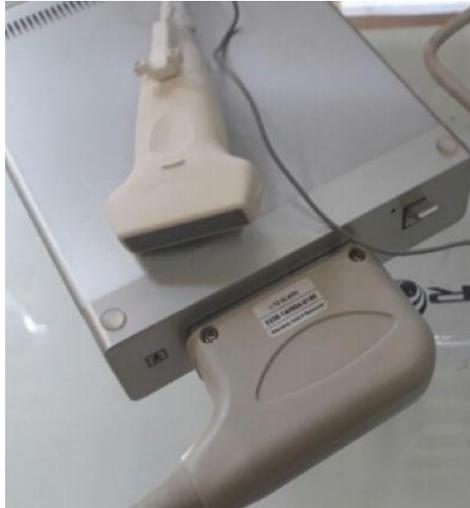

Figure 5. ClarUs EXT-1M ultrasound system and one-dimensional array probe C7-3R50NI-5 for data collection.

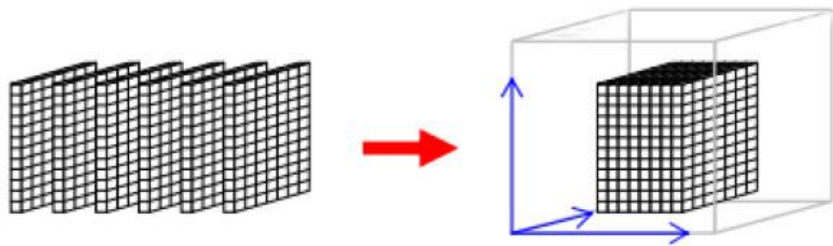

Figure 6. Schematic representation of sequential sampling of images.

For processing and visualization, we need two-dimensional sequential images captured by this system. The images were taken by a radiologist at Imam Khomeini Hospital in Tehran. These liver images were taken from three normal individuals. For the first person, 34 two-dimensional ultrasound images, for the second person, 71 two-dimensional ultrasound images, and for the third person 32 frames were taken. In total, 132 images were recorded and confirmed by *two* radiologists. Figure (7) shows, four 3D data of the liver rendered by TF technique.

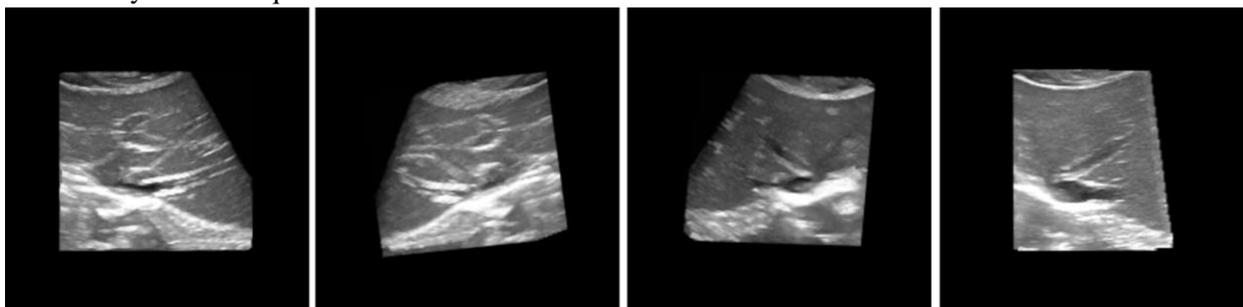

Figure 7. 3D ultrasound images of liver rendered using TF technique.

### 3-2. Preprocessing

As a result of the constructive and destructive interference of recursive waves in the ultrasound imaging system, an unwanted multiplicative point pattern known as speckle noise is formed. The main criteria that should always be observed in speculative noise reduction are:
- reduce noise variance as much as possible in homogeneous areas
- preservation (texture, edges, and lines) of details
- having no artificial side effects

The bilateral filter is a smooth nonlinear filter with the edge retention and noise reduction for images. In this filter, the intensity value of each pixel is replaced by the weighted average of the neighboring pixels. We used the fast bilateral function designed for three-dimensional images in this work [29]. This algorithm is an approximation method that drastically improves speed and is much like its two-dimensional model with some modifications.

### 3-3. Feature extraction

It is one of the steps of the proposed algorithm used is to extract features that can help the physicians to find and visualize areas of interest. Feature extraction is important due to the fact that the visualization of images is performed later through these features. One of the most important features of the edge is the structure of the vessels and the tumor.

3-3-1: Edge detection algorithms in 3D- space

In general, there are edges and axles that have the highest intensities and brightness. In ultrasound images of the liver, these edges and bright spots are the defining borders of veins and tumors. There are different ways to detect the edges. Most of these methods are gradient based, in which we used the Sobel gradient method, which gives us the intensity and direction of the gradient in all 3D-spaces.

**A. Gradient vector flow (GVF)**

The presence of high-level speckle noise in ultrasound images makes segmentation difficult. The image using the contrast agent helps to overcome this. But since the injection has side effects and the surgeon may not be willing to do so, we did not consider the use of contrast material in this work. In this study, we used the GVF technique to isolate vascular and liver tumors presented by Cvancarova et al. [30]. The GVF algorithm is an improved version of the VF algorithm proposed by Xu et al. which has significantly reduced VF noise problems. GVF used snake and active contour methods to detect the edges in the images [31]. Their method also used to detect skeletal and tubular structures. The snake is used as an active (and moving) contour to capture the object boundaries.

**B. Frangi filter**

The vesselness filter was introduced by Frangi et al. [32]. It was originally referred to as a vessel enhancement and amplification filter because the filter can reduce unwanted white noise in the image while maintaining the structure of the vessels. The important step in this algorithm is to compute the Hessian filter matrix from the image. This filter can also obtain the main vessel orientation by calculating the Hessian matrix eigenvalue analysis. To obtain the vascular structure using this algorithm, we used a function that uses the Hessian special vector to calculate similarity in the images to calculate areas containing veins and edges of the image [32].

### 3-4. Visualization

Volumetric visualization is a special field of visualization that displays volumetric information in two-dimensional images on square pages. Volumetric data is a set of samples (called voxels), representing different values in 3D space. Direct volume rendering (DVR) is one of the most important and common methods of visualization, the classification of which is done using the TF. The TF is applied in various forms: one-dimensional and multi-dimensional. The 1D-TF is not efficient for ultrasound images and cannot display details and does not give a good visual view to the user.

The multi-dimensional transfer functions (MDTF) calculates multiple components in the image and provides better image quality but has complex implementations. The method is usually complex, prone to frustration of the user and need to have skills. Hence, many physicians do not prefer to use this method. Instead of using 3D images, they used two-dimensional images and imagined three-dimensional images in their minds. The TF methods are only found in articles and are in little practice. Also, the existing MDTFs cannot be implemented on ultrasound images. The purpose of this study is to eliminate the challenges of TF in the visualization classification stage so that it can be used in the operating room for liver surgery and the surgeon can skillfully and easily receive the 3D image of better quality.

To do this, first we need to find the features that will help us to find areas of interest. In this study, these regions are the vessels and the tumors in the liver. Using the feature extraction methods described above, we obtain the desired features (vascular and tumor structure and tumor and vascular edges) using GVF, Frangi, and gradient algorithms such as Sobel.

The method is to get one degree of importance and one degree of color for each voxel. First, we applied the algorithms to find the properties of the different voxels on the image and obtain features. These features can be binary or normalized between zero and one (divide by Maximum). In this study, we used continuous values (not binary), hence, we will be using all points. The information may be lost due to thresholding of points.

### 3-5. Combining the features

For each attribute, a degree of significance must be chosen between 0 and 1 which must be either chosen by the user relative to the application in question or set in the algorithm by default. These values must follow the formula below, that is, the sum of all degrees of significance must be 1. By user or by default we assign each of these attributes a value between zero and one so that for all attributes this relation exists. If every important function is denoted as K and n is the number of features, then:

$$\sum_{i=1}^{n} K_j = 1 \qquad (1)$$

So automatically the more important a feature becomes the importance of others decreases, and vice versa. The greater the degree of importance, the more important feature becomes for the user. These features are the vessels and the edges of the vessels and tumor. But this algorithm is applicable to all tissues and organs of the body. For example, if there are bones along the artery, another feature could be bone tissue which can be adjusted for presentation. That is why achieving the results is more satisfying and easier than the usual one-dimensional and 2D-TF because the whole process and its selected parameters are performed instantly. They are more tangible to the user and there is an interaction between the user and the system. The user can receive feedback at any time by applying the desired values and change the values if desired. Also, the parameter space is greatly reduced due to the normalization of the values.

Now for each voxel in the image, we must obtain a degree of general significance and a color degree follows the formula (2) and (3) [33]:

$$k(s) = \frac{\sum_{j=1}^{n} X_j(s) I(k_j)}{\sum_{j=1}^{n} X_j(s)} \tag{2}$$

$$I(K_j) = (n . K_j)^2 \quad \Rightarrow \quad k(s) = \frac{\sum_{j=1}^{n} X_j(s) . (n . k_j)^2}{\sum_{j=1}^{n} X_j(s)} \tag{3}$$

This formula is extracted from Berge's work [33] whose parameters are defined as follows:

n: Number of features

k: The degree of importance of each attribute that the user determines

X: Voxels values for each attribute

In the reference paper, X is zero or one, but as stated in this study, we used the results directly to maintain all values and constants. To specify the color, the user defines a color for each feature in the Hue, Saturation, Lightness (HSL) system. The color for each voxel is given in formulas (4) and (5):

$$\delta^{(s)}(s) = \frac{1}{\sum_{j=1}^{n} \omega_j(s)} \cdot \sum_{j=1}^{n} \omega_j(s) . \delta_j^{(s)}. \tag{4}$$

$$\delta^{(H)}(s) = \frac{1}{\sum_{j=1}^{n} \omega_j(s) \delta_j^{(s)}} \cdot \sum_{j=1}^{n} \omega_j(s) . \delta_j^{(s)} . \delta_j^{(H)} \tag{5}$$

and

$$\omega_j(s) = X_j(s) . (n . k_j)^2 \tag{6}$$

where $\delta^{(s)}(s)$ represents saturation and $\delta^{(H)}(s)$ indicates the values of hue in HSL color system. After that these values are converted to RGB space and are applied to all voxels. This formula is also extracted from Berge et al. [33].

Opacity is the most important visual feature in volume rendering which determines the transparency of each voxel in the image. The degree of opacity can be simply increased or decreased by the intensity and gradient. In methods based on the TF, the degree of opacity of each voxel is determined by the TF. In this way, we derived this criterion using the degree of importance obtained in the previous section (formula 7).

$$O_e = O_d(w) . \big(1 + \log(nf_j(w) + 1)\big) . K \tag{7}$$

where, nf is the significance function (k (s)) obtained in the preceding section, and $O_d(w)$ is the initial opacity for each voxel, which is by default adjusted by the intensity of each voxel ($int(v)$) and its gradient

($gm(v)$) as can be seen in formula 8. Therefore, these values (intensity and gradient) are in between zero and one and are normalized.

$$od\ \alpha\ int\ (v).gm(v)\ \rightarrow\ \in [0,1] \qquad (8)$$

This formula is also extracted from Lile's et al. work [10]. The value K is set by the user to interact if desired. If the user prefers that the image remain unchanged and K = 0, the image will not recover. The default value for K is 1. Also if nf = 0, no changes occur in the image.

## 4. Results

As mentioned in the previous section, the first step is speckle noise filtering, which affects the image quality. Table 1 shows the values corresponding to signal-to-noise criterion and mean error criterion of the 3D images.

*Table 1: Filter evaluation values*

| Parameters | Image 1 | Image 2 | Image 3 |
|---|---|---|---|
| MSE | 0.0016 | 0.0014 | 0.0015 |
| PSNR | 27.92 | 28.43 | 28.35 |

The obtained values indicate that this filter is a desirable filter because the noise-peak signal criterion has large values and the mean error criterion has very small values which indicate the desirability of the filter used. Also, if subjectively examined, it can be seen in Figure 8 that the speckle noise is desirably eliminated but does not affect the edges. Since our goal in this study is to enhance visual quality, visual representation of results is significant. We used maximum intensity projection (MIP) visualization to compare which is a usual method for 3D data. It projects the voxels with the highest attenuation value on the volume in the 2D image [34].

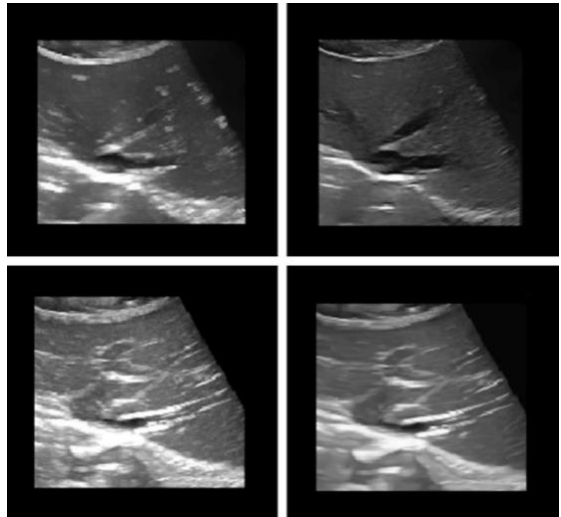

Figure 8. Visualization of 3D images using MIP method: Filtered (left) and unfiltered (right).

The results of the algorithms used to find the edges, and properties and compare them with illustrations using MIP visualization are shown in figures 9, 10, and 11.
Figure 9 illustrates the application of GVF method on the image and Figure 10 employed Sobel gradient method and Figure 11 illustrates the application of Frangi method to the images in MIP visualization.

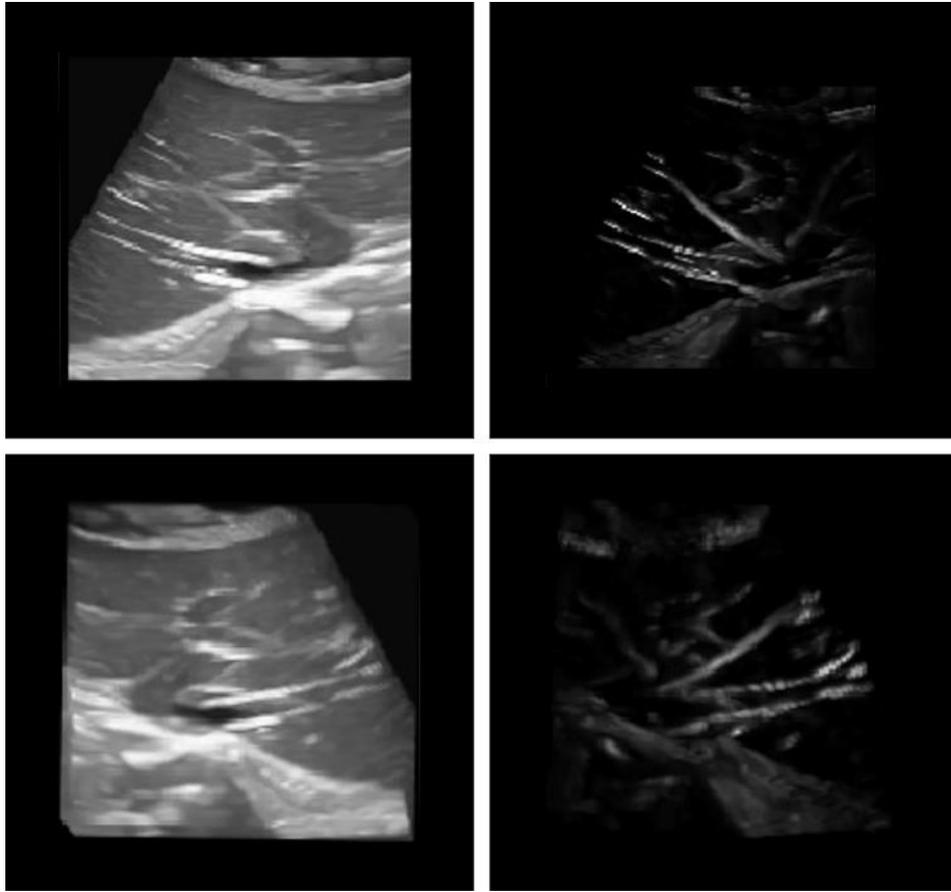

Figure 9. Visualization of images using MIP method: before applying the GVF method (left) after applying the GVF method (right).

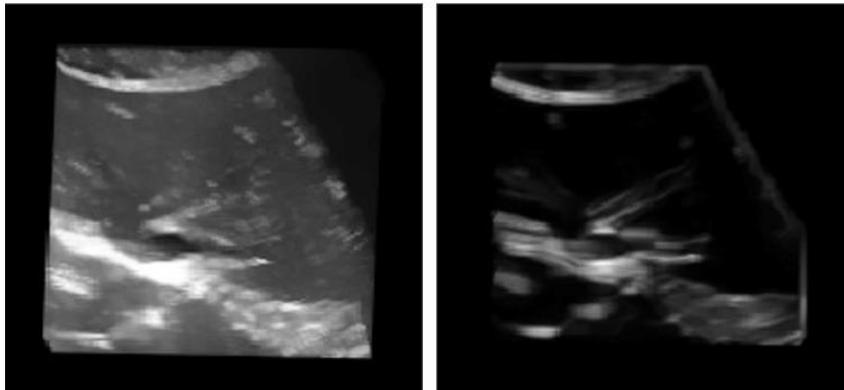

Figure10. Visualization of image before applying the Sobel gradient method (left) and after applying the Sobel gradient method (right).

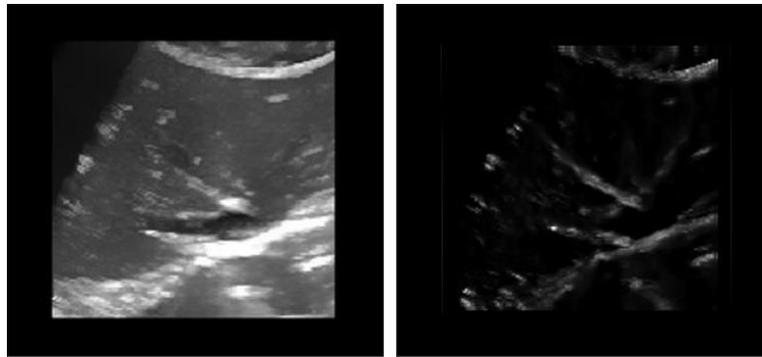

Figure 11. Applying the Frangi method (right) to 3D images using MIP visualization of data (left)

The results of our proposed algorithm is visually acceptable for MIP visualization. After selecting the features and importance degrees by the user, the algorithm is implemented, and the images are visualized with new opacity and color for every voxel. Figure 12 shows the results of gray-scale and colored images illustrated in the new method.

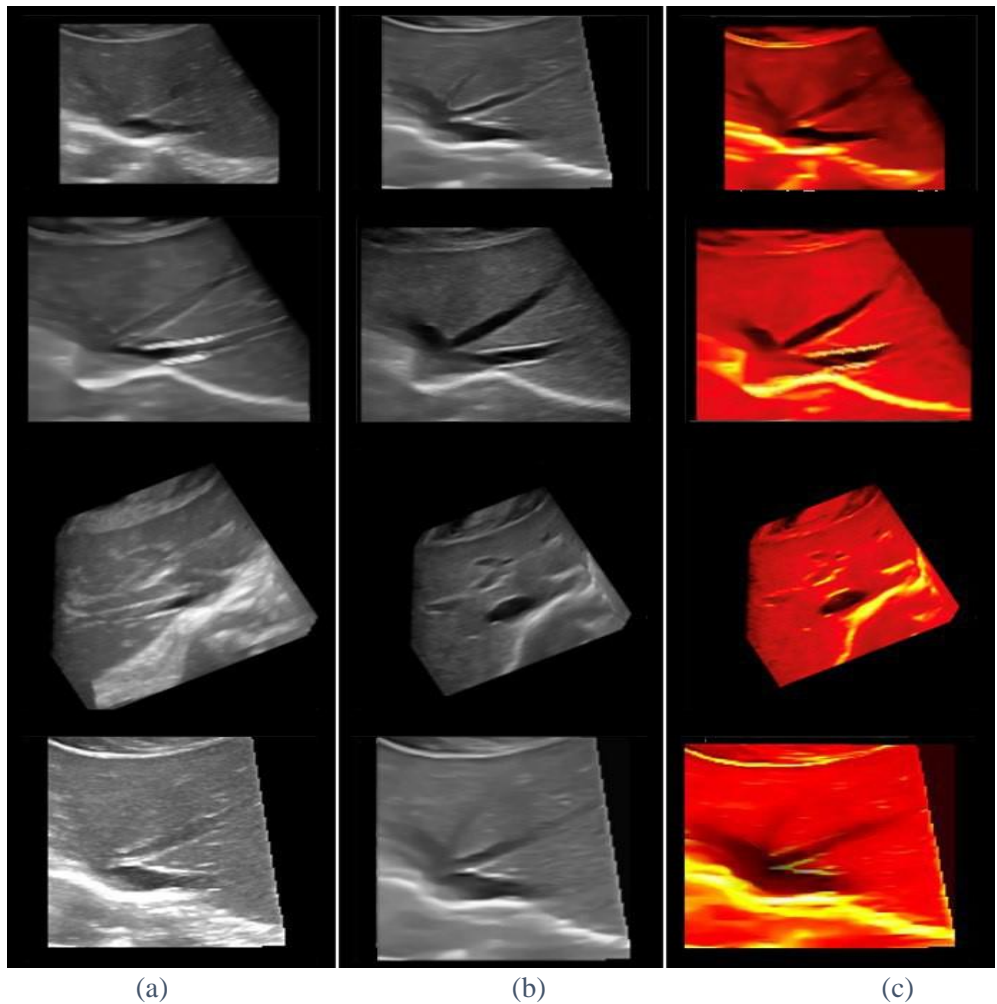

(a)          (b)          (c)

Figure 12. Visualization results using MIP method: proposed (gray) algorithm (a), visualization results (b) and colored image (c).

Figure 12 shows the results obtained by multiplying all the properties. Now the visualization is shown for our features one by one (importance factor for other features is zero) to display the effects of each feature in our visualization.

Figure 13 shows the results obtained by applying MIP visualization to the images (first image), applying the coefficients to only one feature (GVF, Frangi, or Sobel gradient) in our proposed algorithm to show the effect of each feature (images in the middle row), and applying different coefficients to all three features in the liver images (last image). In this figure, different regions are shown in red.

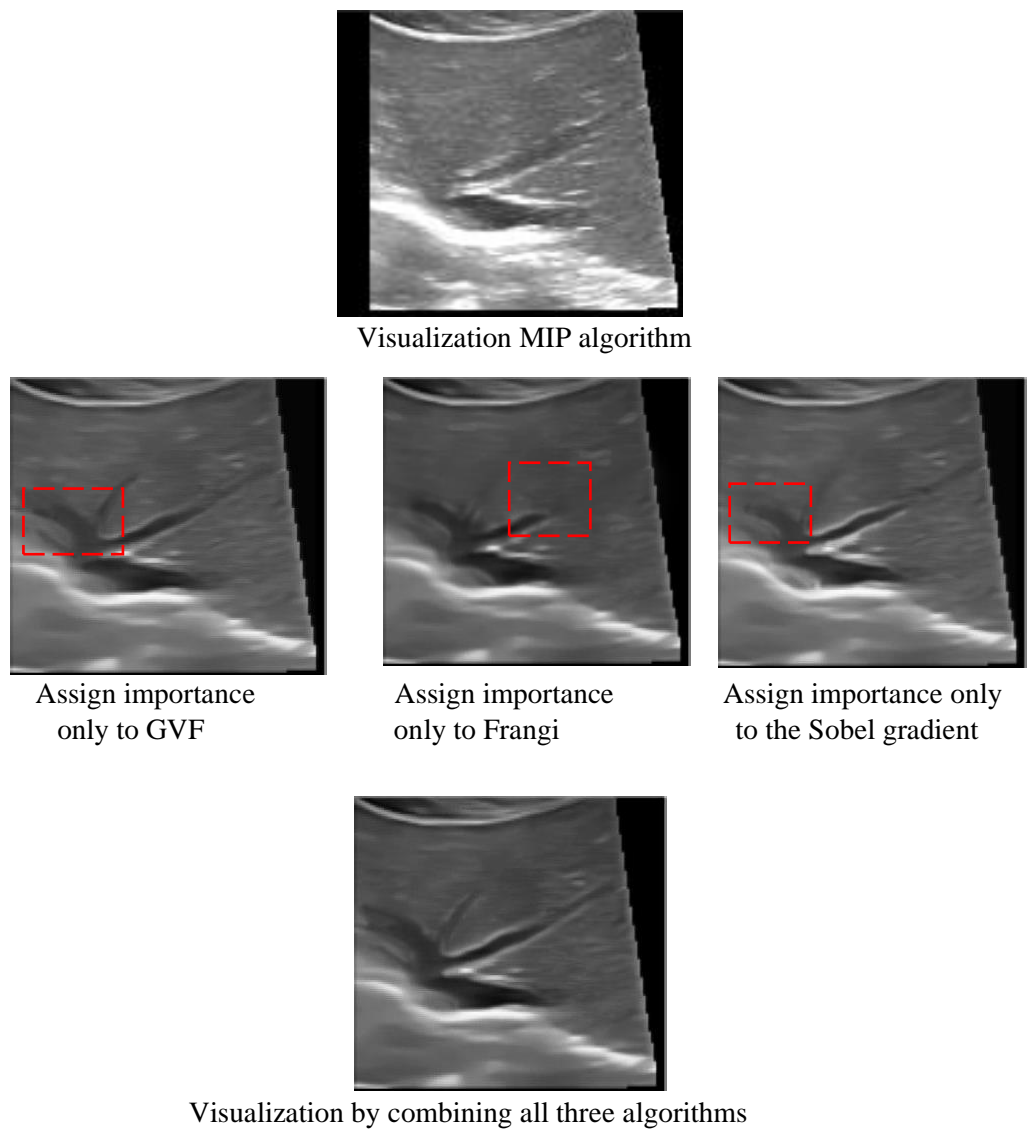

Figure 13. Applying MIP visualization methods and the proposed algorithm with different coefficients in the second case

The algorithms are executed by graphical user interface (GUI) in MATLAB software and displayed in Figure 14 so that the users can view the 3D image in different directions (x) and (y) and (z) and also they can select the type of view with which they want to see the image.

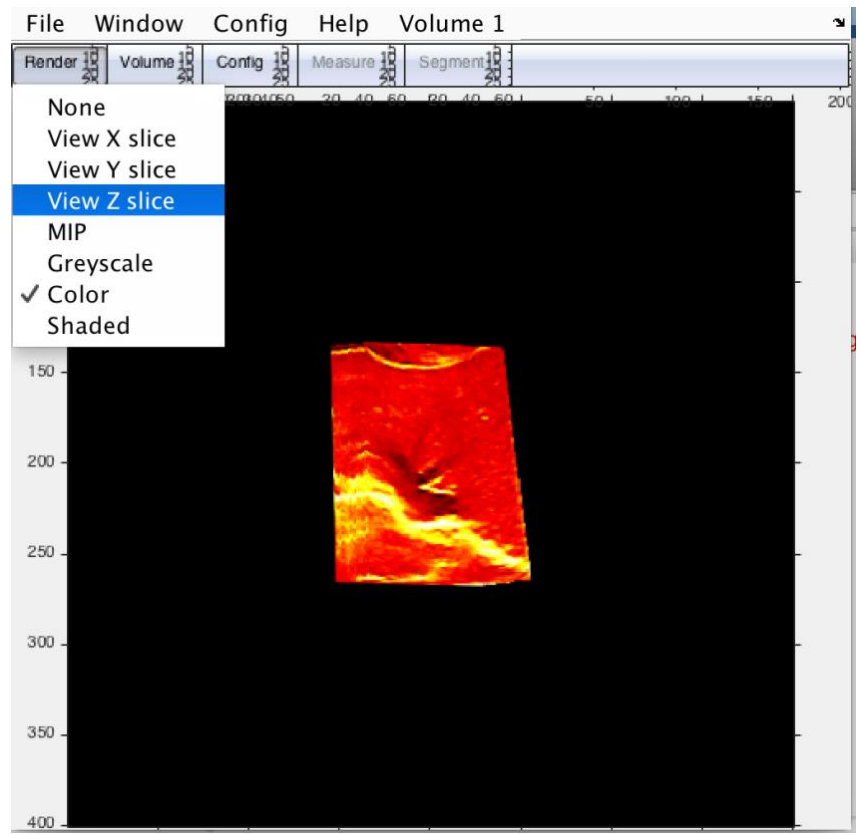

Figure 14 Snapshot of the result obtained using our algorithm.

## 5. Discussion

In this section, we have compared our method of visualization with the MIP method to show its effectiveness. Most existing methods did not use ultrasound data [5]. One of the most common methods employed for visualization is done using TF. Our technique does not has any occlusion effect which exists in 1-D TF methods and effortless unlike MDTFs (which have high dimensions parameters and clinicians find it tedious to use them and are reluctant to use visualizations that use these methods [11]) and surgeons can use it in operation theatre without any previous training.

One of the striking features of this research is that it is one of the first works on visualization of vessels in the liver in 3D ultrasound images. Previous researches do not include challenges and features of liver and vessels [5]. They discussed general ultrasound data but we set all features and methods for better display of liver and its vessels or used simulated data instead of using real ultrasound data. They used other data such as CT or MRI to construct 3D image from them [35]. Few of them focused on registration CT or MRI to ultrasound images for better display [36, 37]. In this work, real 3D ultrasound data of liver and vessels of patients have been obtained and visualization does not need any other extra data (such as CT, and MRI). So it can be used in the operation room and surgeons are more willing to use it since our method is simple unlike MDTF methods, due to the limited access to patients, and tumor images have not been recorded. But the algorithm is designed to respond to tumor images as well. Another limitation is the presence of the shadow of ribs on the images may affect the performance of the algorithm.

## 6. Conclusion and Future Works

Visualization of liver vessels play a vital role in the operation theatre and help to examine the abnormalities in the liver vessels clearly. In this work, we have introduced a novel method to visualize the 3D images of the liver in which the features of the image are used to improve the quality of display. We have made, an attempt to perform the classification of the visualization algorithm, which in most studies used the TF algorithm. The TF is divided into two main categories: one-dimensional and multi-dimensional. The one-dimensional function does not give the desired results to the user and the multi-dimensional function has certain complexities, hence it is not easy for physicians to use. Therefore, the aim of this study is to eliminate the shortcomings of this method. We have obtained different features for every voxel then combined all these features and opacity parameters of each voxel. We have also included a parameter in this formula for the user or surgeons to change these values if they are not satisfied with the created image. Also, each voxel is assigned a color and applied to the image according to its intensity. The user can easily select the colorless and colored image. It also has the ability to view and rotate the 3D image at any view. Few surgeons may be interested in this facility to see the locations of each voxel in relation to other voxels. The proposed method can provide better quality three-dimensional image with more detailed information. Besides, clinicians do not require any additional training to use this model, and also the surgeon's need for radiologists is significantly reduced. Thus, this method is more effective compared with the existing state-of-art methods.

In the future, this method can be applied to three-dimensional images of other organs of the body. It is also possible to completely extract the vascular structure and eliminate the background. Another activity could be to add coordinates of each point in 3D space (which the user wants to access). We also intend to use deep learning techniques to visualize the various organs of the body with higher accuracy and clarity.